# DiffNodesets: An Efficient Structure for Fast Mining Frequent Itemsets

Zhi-Hong Deng[*]

Key Laboratory of Machine Perception (Ministry of Education), School of Electronics Engineering and Computer Science, Peking University, Beijing 100871, China

[*]Corresponding author, email: zhdeng@cis.pku.edu.cn

**Abstract:** Mining frequent itemsets is an essential problem in data mining and plays an important role in many data mining applications. In recent years, some itemset representations based on node sets have been proposed, which have shown to be very efficient for mining frequent itemsets. In this paper, we propose DiffNodeset, a novel and more efficient itemset representation, for mining frequent itemsets. Based on the DiffNodeset structure, we present an efficient algorithm, named dFIN, to mining frequent itemsets. To achieve high efficiency, dFIN finds frequent itemsets using a set-enumeration tree with a hybrid search strategy and directly enumerates frequent itemsets without candidate generation under some case. For evaluating the performance of dFIN, we have conduct extensive experiments to compare it against with existing leading algorithms on a variety of real and synthetic datasets. The experimental results show that dFIN is significantly faster than these leading algorithms.

## 1. Introduction

Frequent itemset mining, first proposed by Agrawal et al. [1], is a fundamental and essential task in the field of data mining because it has been widely used in many important data mining applications. These applications include the discovery of association rules, sequential rules, correlations, episodes, and etc. Since the first proposal of frequent itemset mining, hundreds of algorithms have been proposed on various kinds of extensions and applications, ranging from scalable data mining methodologies, to handling a wide diversity of data types, various extended mining tasks, and a variety of new applications [16].

In recent years, we have presented three kinds of data structure, named Node-list [12], N-list [13], and Nodeset [10], to enhance the efficiency of mining frequent itemset. They are all based on node sets originated from a prefix tree with encoded nodes. The prefix tree employed by Node-list and N-list uses pre-order number and post-order number to encode each node. The only difference between Node-list and N-list is that Node-list use descendant nodes to represent an itemset while N-list represent an itemset by ancestor nodes. Based on Node-list and N-list, two algorithms called PPV [12] and PrePost [13] are proposed respectively for mining frequent itemsets and have shown to be very effective and usually outperform previous algorithms. However, they are memory-consuming because Node-lists and N-lists need to encode a node with pre-order and post-order [10]. In addition, the nodes' encoding model of Node-list and N-list is not suitable to join N-lists (or Node-lists) of two short itemsets to generate the N-list (or Node-list) of a long itemset [10]. To this end, we further present a data structure, namely Nodeset, for mining frequent itemsets. Different from N-list and Node-list, Nodeset requires only the pre-order (or post-order) number of a node to encode the node without the requirement of both pre-order and post-order number. Based on the Nodeset structure, we propose FIN [10] to discover frequent itemsets.

Experiment results show that FIN is comparable with PrePost while it consumes less memory than the latter.

Despite the above advantage of Nodeset, we find the Nodeset cardinality may get very large on some databases. Inspired by the idea of diffset [32], in this paper, we present a novel itemset representation called DiffNodeset, which only keeps track of differences in the Nodesets of a candidate itemset from its generating frequent itemsets. Compared with Nodeset, the cardinality of DiffNodeset is much smaller. Based on DiffNodeset, we design dFIN, an efficient algorithm for mining frequent itemsets. The high efficiency of dFIN is achieved with three techniques: (1) itemsets are represented by DiffNodesets, which is a highly condensed and much smaller structure that can greatly facilitate the mining process of frequent itemsets, (2) it employs a hybrid search strategy to find frequent itemsets in a set-enumeration tree, and (3) it directly enumerates frequent itemsets without candidate generation under some case. An extensive performance study has been conducted to compare the performance of dFIN with FIN, PrePost, FP-growth*, and Eclat_g, which are leading mining algorithms at present. Our extensive experimental study shows that dFIN is efficient and always run faster than other algorithms.

The rest of this paper is organized as follows. In Section 2, we introduce the background and related work for frequent itemset mining. Section 3 introduces some concepts and properties relevant to DiffNodeset. dFIN is described at great length in Section 4. Experiment results are shown in Section 5 and conclusions are given in Section 6.

## 2. RELATED WORK

Formally, the task of frequent itemset mining can be described as follows. Without loss of generality, assume $I = \{i_1, i_2, \ldots, i_m\}$ be the universal item set and $DB = \{T_1, T_2, \ldots, T_n\}$ be a transaction database, where each $T_j$ ($1 \leq j \leq n$) is a transaction which is a set of items such that $T_j \subseteq I$. $P$ is called an itemset if $P$ is a set of items. An itemset containing $k$ items is also called $k$-itemset. Given an itemset $P$, a transaction $T$ is said to contain $P$ if and only if $P \subseteq T$. We define the support of itemset $P$ as the number of transactions in $DB$ that contain $P$. For simplicity, the support of itemset $P$ is denoted as *support*($P$) in this paper. Let $\xi$ be the predefined minimum support threshold and $|DB|$ be the number of transactions in $DB$. We say that itemset $P$ is frequent if its support is not less than $\xi \times |DB|$. Given transaction database $DB$ and threshold $\xi$, the task of mining frequent itemsets is to find the whole set of all itemsets whose supports are not less than $\xi \times |DB|$.

There is much research on algorithm for find all frequent itemsets efficiently. Most of the previously proposed algorithms for mining frequent itemsets can be separated into two classes: candidate generation and pattern growth [5]. Algorithms based on candidate generation first construct candidate itemsets and then identify frequent itemsets from candidate itemsets. These algorithms employ on an anti-monotone property, namely Apriori [2], to prune unpromising itemsets. The Apriori property states that if any $k$-itemset is not frequent, its ($k$ +1) super-itemset also cannot be frequent. The general framework of these algorithms can be described as follows. They generate candidate ($k$+1)-itemsets in the ($k$+1)th pass using frequent $k$-itemsets generated in the previous pass, and count the supports of these candidate itemsets in the database to discover frequent itemsets. A lot of studies, such as [2, 9, 25, 26, 31, 32], belong to the class of candidate generation. The candidate generation method achieves good performance by reducing the size of candidates. However, previous studies reveal that it is highly expensive for candidate generation method to repeatedly scan the database and check a large set of candidates by itemset matching [16].

Different from the candidate generation method, the pattern growth method avoids the need for candidate generation by constructing complex structures that contain sufficient information about frequent itemsets within the database. The FP-growth algorithm, proposed by Han et al. [17], is the classic and fundamental pattern growth algorithm. FP-growth has been shown to be very efficient in the mining of dense databases as the FP-tree structure adopted by FP-growth concisely encapsulates sufficient itemset information and no candidate itemsets are generated. Similar to FP-growth, some studies [15, 17, 21, 22] adopt pattern growth method to mine frequent itemsets. The pattern growth method wins an advantage over the candidate generation method by reducing search space and directly generating frequent itemsets without candidate generation. However, the data structures adopted by pattern growth algorithms, such as FP-tree, are complex [30]. In addition, recurrently building such structures makes pattern growth algorithms inefficient when datasets are sparse [13].

In recent years, we have proposed three kinds of structure for representing itemsets: Node-list [12], N-list [13], and Nodeset [10], to facilitate the mining of frequent itemsets. They are based on a prefix coding tree, which store the sufficient information about frequent itemsets. Node-list and N-list is based on a PPC-tree, which is a prefix tree with each node encoded by its pre-order number and post-order number. The N-list (or Node-list) of an itemsets is a set of nodes in the PPC-tree. The only difference between N-list and Node-list lies in that the Node-list of an itemset consists of descendant nodes while its N-list consists of ancestor nodes. N-lists (or Node-lists) have two important properties. First, the support of an itemset is the sum of counts registering in the nodes of its N-list (or Node-list). Second, the N-list (or Node-list) of a ($k$ +1)-itemset can be constructed by joining the N-lists (or Node-lists) of its subset with length of $k$ with linear computation complexity. Compared with the vertical structures for representing itemsets, such as Diffset [32], the size of N-list or Node-list is much smaller. Compared with FP-tree [17], they are more simple and flexible. Therefore, the algorithms based on N-list or Node-list show great efficient and outperform the existing classic algorithms, such as dEclat and FP-growth. Compared with Node-lists, N-lists have two advantages. The first one is that the length of the N-list of an itemset is much smaller than the length of its Node-list. The other one is that N-lists have property called single path property, which can be employed to directly mining frequent itemsets without generating candidate itemsets in some case. These make that PrePost [13], the mining algorithm based on N-lists, is more efficient than PPV [12], the mining algorithm based on Node-lists. In recent years, we have developed PrePost into PrePost+ [11] by employing a very efficient pruning technique. Although N-list and Node-list are efficient structures for mining frequent itemsets, they need to contain pre-order and post-order number, which is memory-consuming and inconvenient to mine frequent itemsets. Therefore, we further propose Nodeset, which represent itemsets with only pre-order (or post-order). Experiment results on some real and synthetic datasets show that Nodeset is an effective structure and the corresponding algorithm, FIN, performs better than PrePost [10].

In addition, similar structure named NC_set [14] has been proposed to mine erasable itemsets [7], a new kind of mining task that is different from frequent itemsets, and the experimental results show that NC_set is effective and the algorithm based on it is very efficient and far outperforms those previously proposed algorithms [14].

# 3. DiffNodeset: Design and construction

In this section, we first introduce relevant related concepts and then present DiffNodeset.

## 3.1. PPC-tree

Given a database and a minimum support threshold, the PPC-tree [13] is defined as follows.

**Definition 1.** PPC-tree is a tree structure:
   (1) It consists of one root labeled as "null", and a set of item prefix subtrees as the children of the root.
   (2) Each node in the item prefix subtree consists of five fields: *item-name*, *count*, *children-list*, *pre-orde*, and *post-order*. *item-name* registers which item this node represents. *count* registers the number of transactions presented by the portion of the path reaching this node. *children-list* registers all children of the node. *pre-order* is the pre-order number of the node and *post-order* is the post-order number of the node.

Given *DB*, the transaction database, and $\xi$, the given minimum support threshold, the construction of the PPC-tree is described as follows.

*Construct_PPC-tree* (*DB*, $\xi$)
(1) Scan *DB* once to find $F_1$, the set of frequent items.
(2) Sort $F_1$ in support descending order as $L_1$.
(3) Create the root of a PPC-tree, *Tr*, and label it as "null".
(4) For each transaction *T* in DB do
(5)    Delete all infrequent items from *T* and then sort *T* as $T_f$ according to the order of $L_1$. Let the sorted frequent-item list in *Trans* be [*p* | *P*], where *p* is the first element and *P* is the remaining list.
(6) Call *insert_tree*([*p* | *P*], *Tr*).
(7) Scan the PPC-tree to generate the *pre-order* and *post-order* of each node by the pre-order traversal.
(8) Return *Tr* and $L_1$.

Note that, the function *insert_tree*([*p* | *P*], *Tr*) is performed as follows. If *Tr* has a child *N* such that *N.item-name* = *p.item-name*, then increase *N*'s count by 1; else create a new node *N*, with its count initialized to 1, and add it to *Tr*'s *children-list*. If *P* is nonempty, call *insert_tree*(*P*, *N*) recursively.

For better understanding the concept and the construction algorithm of PPC-tree, let's examine the following example.

**Example 1.** Let *DB* be the transaction database as shown in Table 1, and minimum support threshold $\xi$ 0.4. By scanning *DB*, we obtain the set of frequent item set, $F_1$ (={*a, b, c, d, e*}). The right column of Table 1 shows the sorted transactions with deleting infrequent items.

Figure 1 shows the PPC-tree which is constructed from the database shown in Example 1 after running Algorithm 1. The number pair on the outside of a node is its pre-order (left number) and post-order (right number). The letter on the inside of a node is the item registered in it while the number on the inside of the node is the count of all transactions which register the item in the node.

Table 1. An example of transaction database

| ID | Items | Sorted frequent items |
|---|---|---|
| 1 | c, g | c |
| 2 | a, b, c | c, b, a |
| 3 | a, b, c, d, e | e, d, c, b, a |
| 4 | a, b, c, e | e, c, b, a |
| 5 | a, b, c, e, f | e, c, b, a |
| 6 | d, e, f | e, d |
| 7 | d, e, g | e, d |
| 8 | d, e, i | e, d |
| 9 | d, e, i, h | e, d |
| 10 | d, e, g, i | e, d |

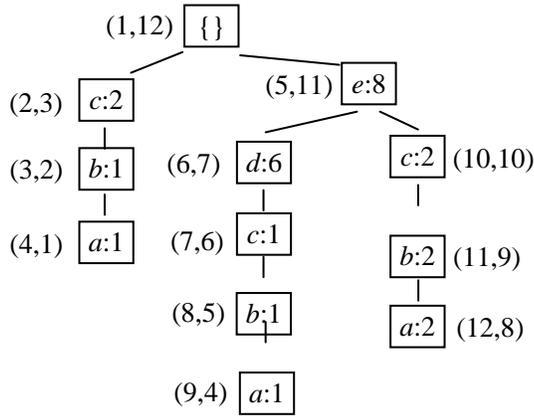

Figure 1. The PPC-tree in Example 1

Given a PPC-tree, we have the following Properties [13].

**Property 1.** For any node $N_1$ and $N_2$ ($N_1 \neq N_2$) in the PPC-tree, $N_1$ is an ancestor of $N_2$ if and only if $N_1$.pre-order $< N_2$.pre-order and $N_1$.post-order $> N_2$.post-order.

**Property 2.** For any node $N_1$ and $N_2$ ($N_1 \neq N_2$) in the PPC-tree, which register the same item ($N_1$.item-name $= N_2$.item-name), if $N_1$.pre-order $< N_2$.pre-order, then $N_1$.post-order $< N_2$.post-order.

### 3.2. Nodesets

**Definition 2.** Let $L_1$ be the ordered set of frequent items sorted in support descending order. For any two items $i_1$ and $i_2$ ($i_1, i_2 \in L_1$), we denote $i_1 \prec i_2$ if and only if $i_2$ is ahead of $i_1$ in $L_1$.

For example, in terms of example 1, we have $a \prec b \prec c \prec d \prec e$.

For the sake of discussion, the following conventions hold in the remainder of this paper:

**Convention 1.** An itemset is considered as ordered, and all the items in the itemset are sorted in $\prec$ order.

For example, the itemset consisting of item *e, a,* and *c* is denoted by *ace*.

**Definition 3. (PP-code)** Given a node $N$ in a PPC-tree, we call triple ($N$.pre-order, $N$.post-order, *count*) the PP-code of $N$.

Since $N$.pre-order uniquely identify node $N$, the PP-code of $N$ can also be defined as ($N$.pre-order, *count*). Obviously, these two definitions is equivalent. In this paper, we use one or the other definition when needed.

**Definition 4. (Nodesets of items)** Given a PPC-tree, the Nodeset of frequent item $i$ is a ordered sequence of all the PP-codes of nodes registering $i$ in the PPC-tree, where PP-codes are sorted in *pre-order* ascendant order.

Figure 2 shows the Nodesets of all frequent items in Example 1.

$$e \longrightarrow \{(5,11, 8)\}$$
$$d \longrightarrow \{(6, 7, 6)\}$$
$$c \longrightarrow \{(2, 3, 2), (7, 6, 1), (10, 10, 2)\}$$
$$b \longrightarrow \{(3, 2, 1), (8, 5, 1), (11, 9, 2)\}$$
$$a \longrightarrow \{(4, 1, 1), (9, 4, 1), (12, 8, 2)\}$$

Figure 2. The Nodesets of frequent items in Example 1

Note that, the definition of items' Nodesets in this paper is the same as N-lists presented in [13] instead of the original definition presented in [10]. The only difference of these two kinds of Nodeset lie in that the one used in this paper contains both pre-order and post-order number while the other contain only pre-order number. However, the difference can be negligible since both pre-order and post-order can uniquely identify a node in a PPC-tree. In fact, the two kinds of Nodeset can replace each other. In this paper, we adopt Definition 3 for the definition of items' Nodesets in order that we can design efficient method to build the DiffNodesets of 2-itemsets.

Based on Property 1 and Property 2, we can infer the following properties.

**Property 3.** Given an item, assume $\{(x_1, y_1, z_1), (x_2, y_2, z_2), …, (x_l, y_l, z_l)\}$ is its Nodeset. We have that $x_1 < x_2 <…< x_l$ and $y_1 < y_2 <…< y_l$.

Proof. According to Definition 4, we have $x_1 < x_2 <…< x_l$. Let $s$ and $t$ ($1 \leq s < t \leq l$) be two indexes. According to Property 2, we have $y_s < y_t$ since $x_s < x_t$. Therefore, Property 3 holds. □

Property 3 indicates that the elements in the Nodeset of an item are sorted in both *pre-order* ascendant order and *post-order* ascendant order.

**Definition 5. (Nodesets of 2-itemsets)** Given items $i_1$ and $i_2$ ($i_1, i_2 \in L_1 \wedge i_1 \prec i_2$), the Nodeset of 2-itemset $i_1i_2$, denoted as Nodesets$_{i_1i_2}$, which is defined as follows:
Nodesets$_{i_1i_2}$ = {($x$.pre-order, $x$.count) | $x \in$ Nodesets$_{i_1} \wedge (\exists\, y \in$ Nodesets$_{i_2}$, the node corresponding to $y$ is an ancestor of the node corresponding to $x$)}
where Nodesets$_{i_1}$ and Nodesets$_{i_2}$ are the Nodesets of item $i_1$ and $i_2$ respectively.

**Definition 6. (Nodesets of $k$-itemsets)** Let $P = i_1i_2…i_k$ be an itemset ($i_j \in L_1$ and $i_1 \prec i_2 \prec i_3 \prec … \prec i_k$). We denote the Nodeset of $P_1(= i_1i_2…i_{k-2}i_{k-1})$ as $Nodeset_{P1}$ and the Nodeset of $P_2(= i_1i_2…i_{k-2}i_k)$ as $Nodeset_{P2}$. The Nodeset of $P$, denoted as $Nodeset_P$, is defined as intersection of $Nodeset_{P1}$ and $Nodeset_{P2}$. That is, $Nodeset_P = Nodeset_{P1} \cap Nodeset_{P2}$.

According to Definition 5, the Nodeset of *ce* is {(7, 1), (10, 2)} since node 5 is an ancestor of node 7 and 10, and is not an ancestor of node 2. Similarly, the Nodeset of *cd* is {(7, 1)}. Based on the above results, we have that the Nodeset of *cde* is {(7, 1)} in terms of Definition 6.

The Nodeset has one important property as follows [10].

**Theorem 1.** Given itemset $P$, we denote its Nodeset as $NS_p$, the support of $P$, $support(P)$, is equal to $\sum_{(E \in NSp)} E.count$.

For example, the support of *ce* is 3, which is equal to 1, the *count* of (7, 1), plus 2, the *count* of (10, 2).

Theorem 1 will be used to proof some properties of DiffNodesets in the following subsection.

### 3.3. DiffNodesets

Based on the concept of Nodeset, we introduce DiffNodeset in this subsection.

**Definition 7. (DiffNodesets of 2-itemsets)** Given item $i_1$ and $i_2$ ($i_1, i_2 \in L_1 \wedge i_1 \prec i_2$), we denote their Nodesets as $Nodesets_{i1}$ and $Nodesets_{i2}$ respectively. The DiffNodeset of 2-itemset $i_1i_2$, denoted as $DiffNodesets_{i1i2}$, is defined as follows:

$DiffNodesets_{i1i2}$ = {($x.pre\text{-}order$, $x.count$) | $x \in Nodesets_{i1} \wedge \neg (\exists y \in Nodesets_{i2}$, the node corresponding to $y$ is an ancestor of the node corresponding to $x$)}.

where $Nodesets_{i1}$ and $Nodesets_{i2}$ are the Nodesets of item $i_1$ and $i_2$ respectively. In addition, the elements in $DiffNodesets_{i1i2}$ are sorted in *pre-order* ascendant order.

As shown in Figure 2, the Nodeset of *e* and *c* are {5, 8} and {(2, 2), (7, 1), (10, 2)}. Let's see how to generate the DiffNodeset of *ce*. From Figure 1, we find node 5 is not an ancestor of node 2, which is the corresponding node of (2, 2). Therefore, we insert (2, 2) into the DiffNodeset of *ce*. Similarly, we find node 5 is an ancestor of node 7 and 10, which are the corresponding node of (7, 1) and (10, 2) respectively. Thus, (7, 1) and (10, 2) are not inserted into the DiffNodeset of *ce*. Up to now, all elements in the Nodeset of *c* are checked. Therefore, the DiffNodeset of *ce* is {(2, 2)}. Similarly, the DiffNodeset of *cd* is {(2, 2), (10, 2)}.

Given item item $i_1$ and $i_2$, we assume the length of their Nodesets is $m$ and $n$ respectively. We can build the $DiffNodesets_{i1i2}$ by checking the ancestor-descendant relation between each element in $i_1$'s Nodeset and each element in $i_2$'s Nodeset. This naïve method is clearly inefficient since its computational complexity is O($m*n$). By employing Property 1 and Property 3, we propose ***Build_2-itemset_DN*()**, a method with linear complexity.

***Build_2-itemset_DN*()** compares the elements (PP-codes) from the beginning of two Nodesets until all elements of one Nodeset are finished as shown by Line (4). For the two elements in comparison, there are three cases. Case one is $N_x[k].post\text{-}order > N_y[j].post\text{-}order$ shown by Line (5). In this case, the node corresponding to $N_y[j]$ is not an ancestor of the node corresponding to $N_x[k]$ in terms of Property 1. In addition, the Nodesets of items are sorted in *post-order* ascendant order in terms of Property 3. Therefore, the node corresponding to $N_y[j]$ can also not be an ancestor of the nodes corresponding to the elements which are listed after $N_x[k]$ in the Nodeset of $i_x$. Therefore, $N_y[j]$ is no longer considered and the next element to $N_y[j]$ is selected to compare with $N_x[k]$, as shown by Line (5) and (6). Case two is $N_x[k].post\text{-}order < N_y[j].post\text{-}order$ and $N_x[k].pre\text{-}order > N_y[j].pre\text{-}order$. This case means that the node corresponding to $N_y[j]$ is an ancestor of the node corresponding to $N_x[k]$ in terms of Property 1. According to Definition 7, $N_x[k]$ is undesirable. Therefore, the next element to $N_x[k]$ is selected to perform comparison, as shown by Line (9). The last case is $N_x[k].post\text{-}order < N_y[j].post\text{-}order$ and $N_x[k].pre\text{-}order < N_y[j].pre\text{-}order$ as shown by Line (10). This case means that the node corresponding to $N_y[j]$ is not an ancestor of the node corresponding to $N_x[k]$ in terms of Property 1. In addition, according to Definition 4, the Nodesets of items are sorted in *pre-order* ascendant order. Therefore, in this case, the node corresponding to $N_x[k]$ can also not be a descendant of the nodes corresponding to the elements which are listed after $N_y[j]$ in the Nodeset of $i_y$. That is, $N_x[k]$ is desirable. Thus, a corresponding result is generated and appended to the DiffNodeset of $i_x i_y$, as shown by Line (11). Meanwhile, the next element to $N_x[k]$ is selected to perform comparison as shown by Line (12). When all comparison operations are finished, we should check whether there exist some elements in the Nodeset of $i_x$, which do not take part in the comparison operation. Clearly, these elements are also desirable according to Definition 7. Therefore, all results corresponding to these elements are generated and appended to the DiffNodeset of $i_x i_y$, as shown by Line (16) to (21). Finally, **Build_2-itemset_DN**() outputs the DiffNodeset of $i_x i_y$.

In fact, **Build_2-itemset_DN**() adopts the strategy of 2-way comparison to build the DiffNodeset of $i_x i_y$. According to the principle of 2-way comparison, the computational complexity of **Build_2-itemset_DN**() is $O(m+n)$, where $m$ and $n$ are the lengths of corresponding Nodesets.

**Procedure Build_2-itemset_DN($i_x i_y$)**
(1) $DN_{xy} \leftarrow \emptyset$;
(2) $k \leftarrow 0$ and $j \leftarrow 0$;
(3) $l_x \leftarrow$ the length of $N_x$ (Nodeset of $i_x$) and $l_y \leftarrow$ the length of $N_y$ (Nodeset of $i_y$);
(4) **While** $k < l_x \wedge j < l_y$ **do**
(5)    **If** $N_x[k].post\text{-}order > N_y[j].post\text{-}order$ **then**
(6)       $j \leftarrow j + 1$;
(7)    **Else**
(8)       **If** $N_x[k].post\text{-}order < N_y[j].post\text{-}order$ and $N_x[k].pre\text{-}order > N_y[j].pre\text{-}order$ **then**
(9)          $k \leftarrow k + 1$;
(10)       **Else**
(11)          $DN_{xy} \leftarrow DN_{xy} \cup \{(N_x[k].post\text{-}order, N_x[k].count)\}$;
(12)          $k \leftarrow k + 1$;
(13)       **Endif**
(14)    **Endif**
(15) **Endwhile**
(16) **If** $k < l_x$ **then**
(17)    **While** $k < l_x$ **do**
(18)       $DN_{xy} \leftarrow DN_{xy} \cup \{(N_x[k].post\text{-}order, N_x[k].count)\}$;
(19)       $k \leftarrow k + 1$;
(20)    **Endwhile**

(21) **Endif**
(22) **Return** $DN_{xy}$.

The DiffNodeset of a 2-itemset has the following property.

**Theorem 2.** Given 2-itemset $i_1i_2$, we denote its DiffNodeset as $DN_{12}$. The support of $i_1i_2$, $support(i_1i_2)$, is equal to $support(i_1) - \sum_{(E \in DN12)} E.count$.

**Proof.** We denote the Nodeset of $i_1$ as $NS_1$ and the Nodeset of $i_1i_2$ as $NS_{12}$. According to Definition 5 and 7, we have that
$$NS_1 = NS_{12} \cup DN_{12}, \quad (1)$$
$$NS_{12} \cap DN_{12} = \emptyset. \quad (2)$$

According to Theorem 1, we have that

$support(i_1) = \sum_{(E \in NS1)} E.count$ \hfill (Theorem 1)

$= \sum_{(E \in NS12 \cup DN12)} E.count$ \hfill (using Eq. (1))

$= \sum_{(E \in NS12)} E.count + \sum_{(E \in DN12)} E.count$ \hfill (using Eq. (2))

$= support(i_1i_2) + \sum_{(E \in DN12)} E.count$ \hfill (Theorem 1)

Therefore, Theorem 2 holds. □

For example, the *support* of *ce* is 3, which is equal to 5, the support of *c*, minus 2, the sum of all counts in the DiffNodeset of *ce*.

**Definition 8. (DiffNodesets of $k$-itemsets)** Let $P = i_1i_2i_3\ldots i_k$ be an itemset ($i_j \in L_1$ and $i_1 \prec i_2 \prec i_3 \prec \ldots \prec i_k$). We denote the Nodeset of $P_1 = i_1i_2\ldots i_{k-2}i_{k-1}$ as Nodeset$_{P1}$ and the Nodeset of $P_2 = i_1i_2\ldots i_{k-2}i_k$ as Nodeset$_{P2}$. The DiffNodeset of $P$, denoted as DiffNodeset$_P$, is defined as follows:
$$\text{DiffNodeset}_P = \text{Nodeset}_{P1} / \text{Nodeset}_{P2}.$$
where "/" is the operation of set difference.

For example, the Nodesets of *ce* and *cd* are {(7:1), (10:2)} and {(7:1)} respectively. In terms of Definition 8, the DiffNodeset of *cde* is null (= {{(7:1)} / {(7:1), (10:2)} = ∅).

For the DiffNodeset of a $k$-itemset, The following Theorem 3 holds.

**Theorem 3.** Given itemset $P = i_1i_2i_3\ldots i_k$ and $P_1 = i_1i_2\ldots i_{k-2}i_{k-1}$, we denote the DiffNodeset of $P$ as $DN_p$. The support of $P$, $support(P)$, can be computed as follows:
$$support(P) = support(P_1) - \sum_{(E \in DNp)} E.count.$$

**Proof.** Let $P_2$ be itemset $i_1i_2\ldots i_{k-2}i_k$. We denote the Nodeset of $P$, $P_1$, and $P_2$ as $NS_p$, $NS_1$, and $NS_2$ respectively. According to Definition 6, we have $NS_p = NS_1 \cap NS_2$. According to Definition 8, we have $DN_p = NS_1 / NS_2$. Therefore, we have that
$$NS_1 = (NS_1 \cap NS_2) \cup (NS_1 / NS_2) = NS_p \cup DN_p, \quad (3)$$
$$NS_p \cap DN_p = \emptyset. \quad (4)$$

According to Theorem 1, we have that

$support(P_1) = \sum_{(E \in NS1)} E.count$ \hfill (Theorem 1)

$= \sum_{(E \in NSp \cup DNp)} E.count$ \hfill (using Eq. (3))

$= \sum_{(E \in NSp)} E.count + \sum_{(E \in DNp)} E.count$ \hfill (using Eq. (4))

$$= support(P) + \sum_{(E \in DN_P)} E.count \quad \text{(Theorem 1)}$$
Therefore, Theorem 2 holds. □

For example, the support of *cde* is equal to 1, which is the result of 1 (the support of *cd*) minus 0 (the sum of all counts in the DiffNodeset of *cde*, which is null).

DiffNodeset has an important property as follows, which makes that DiffNodesets can be directly computed without Nodesets.

**Theorem 4.** Let $P = i_1 i_2 i_3 \ldots i_k$ be an itemset ($i_j \in L_1$ and $i_1 \prec i_2 \prec i_3 \prec \ldots \prec i_k$). We denote the DiffNodeset of $P_1 = i_1 i_2 \ldots i_{k-2} i_{k-1}$ as $DN_1$ and the DiffNodeset of $P_2 = i_1 i_2 \ldots i_{k-2} i_k$ as $DN_2$. The DiffNodeset of $P$, $DN_P$, can be computed as follows:
$$DN_P = DN_2 \,/\, DN_1.$$

**Proof.** Let's denote item $i_1 i_2 \ldots i_{k-3} i_{k-2}$ and $i_1 i_2 \ldots i_{k-3} i_{k-1}$ by $X$ and $Y$ respectively. For simplicity, we denote the Nodesets of itemset $X$, $Y$, $P_1$, and $P_2$ by $NS_X$, $NS_Y$, $NS_1$, and $NS_2$ respectively. According to Definition 6 and Definition 8, we have that

$$NS_1 = NS_X \cap NS_Y, \quad (5) \quad \text{(Definition 6)}$$
$$DN_1 = NS_X \,/\, NS_Y. \quad (6) \quad \text{(Definition 8)}$$

Based on Eq. (5) and (6), we have that
$$NS_X = (NS_X \cap NS_Y) \cup (NS_X \,/\, NS_Y) = NS_1 \cup DN_1, \quad (7)$$
$$NS_1 \cap DN_1 = \emptyset. \quad (8)$$

Therefore, we have that
$$NS_1 = NS_X \,/\, DN_1. \quad (9)$$

By the same way, we have that
$$NS_X = NS_2 \cup DN_2, \quad (10)$$
$$NS_2 \cap DN_2 = \emptyset, \quad (11)$$
$$NS_2 = NS_X \,/\, DN_2. \quad (12)$$

Finally,
$$\begin{aligned}
DN_P &= NS_1 \,/\, NS_2 && \text{(Definition 8)} \\
&= (NS_X \,/\, DN_1) \,/\, (NS_X \,/\, DN_2) && \text{(using Eq. (5) and (8))} \\
&= NS_X \cap (DN_1)^T \cap DN_2 && ((DN_1)^T \text{ means the complement of A}) \\
&= DN_2 \cap (DN_1)^T && \text{(using Eq. (10) to infer } DN_2 \subseteq NS_X) \\
&= DN_2 \,/\, DN_1. \quad \square
\end{aligned}$$

For example, we know that the DiffNodesets of *ce* and *cd* are {(2, 2)} and {(2, 2), (10, 2)}. Therefore, DiffNodesets of *cde* can be directly computed by {(2, 2)} / {(2, 2), (10, 2)} (= ∅).

## 4. FIN+: THE PROPOSED METHOD

The framework of FIN+ consists of three sequential parts. The first one is to construct the PPC-tree and identify all frequent 1-itemsets with their Nodesets. The second one is to scan the PPC-tree to find all frequent 2-itemsets and their DiffNodesets; (3) mine all frequent k(>2)-itemsets. For the sake of mining efficiency, FIN+ employs two techniques adopted by FIN, which are set-enumeration tree [24] and superset equivalence property [11].

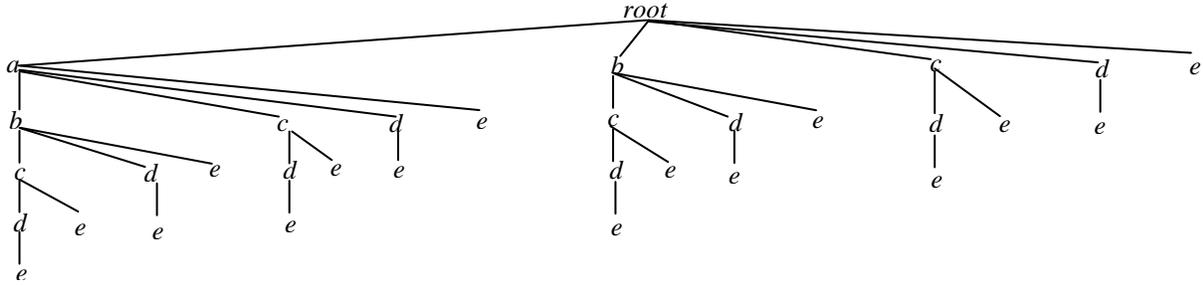

Figure 3. An example of set-enumeration tree

FIN+ uses a set-enumeration tree to represent the search space. Given a set of items $I = \{i_1, i_2, \ldots, i_m\}$ where $i_1 \prec i_2 \prec \ldots \prec i_m$, a set-enumeration tree can be constructed as follows [8]. Firstly, the root of the tree is created. Secondly, the $m$ child nodes of the root registering and representing $m$ 1-itemsets are created, respectively. Thirdly, for a node representing itemset $\{i_{js} i_{js-1}\ldots i_{j1}\}$ and registering $i_{js}$, the $(m - j_s)$ child nodes of the node representing itemsets $\{i_{js+1}i_{js}i_{js-1}\ldots i_{j1}\}$, $\{i_{js+2}i_{js}i_{js-1}\ldots i_{j1}\}$,…, $\{i_m i_{js}i_{js-1}\ldots i_{j1}\}$ and registering $i_{js+1}, i_{js+2},\ldots, i_m$ respectively are created. Finally, the set-enumeration tree is built by executing the third step repeatedly until all leaf nodes are created. The set-enumeration tree represented in Figure 3 is the search space for Example 1. In Figure 3, the bottom left node represents itemset *abcde* and registers item *e*.

Algorithm 1 shows the pseudo-code of FIN+. Line (1) initializes $F$, which is used to store frequent itemsets, by setting it to be null. Line (2) constructs the PPC-tree and finds $F_1$, the set of all frequent 1-itemset, by calling procedure ***Construct_PPC-tree*** (). Line (4) and (5) generate the Nodeset of each frequent item by scanning the PPC-tree. Line (8) calls procedure ***Build_2-itemset_DN***() to generate the DiffNodeset of each 2-itemset. Line (9) computes the support of each 2-itemset according to Theorem 2. Line (10) and (11) check whether $i_x i_y$ is frequent. Line (15) – (17) generate all frequent $k$-itemsets ($k \geq 3$) by calling procedure ***Constructing_Pattern_Tree*** () to generate all frequent $k$-itemsets ($k \geq 3$) extended from frequent 2-itemsets.

**Algorithm 1: (FIN+ Algorithm)**
**Input:** A transaction database *DB* and a minimum support $\xi$.
**Output:** $F$, the set of all frequent itemsets.
(1)  $F \leftarrow \varnothing$;
(2)  **Call *Construct_PPC-tree*** (*DB*, $\xi$) to construct the PPC-tree and find $F_1$, the set of all frequent items;
(3)  $F \leftarrow F \cup F_1$;
(4)  Traverse the PPC-tree with pre-order, For each node $N$ in the PPC-tree do
(5)    Append ($N.pre\text{-}order$, $N.count$) to the Nodeset of item $N.\,item\text{-}name$;
(6)  $F_2 \leftarrow \varnothing$;
(7)  **For** each 2-itemset $i_x i_y$ **do**
(8)    **Call *Build_2-itemset_DN*($i_x i_y$)** to generate its DiffNodesets, $DN_{xy}$;
(9)    $i_x i_y.support \leftarrow i_x.support - \sum_{(E \in DN_{xy})} E.count$;
(10) **If** $i_x i_y.support < \xi \times |DB|$, **then**
(11)    $F_2 \leftarrow F_2 \cup \{i_x i_y\}$;
(12) **Endif**
(13) **Endfor**

(14) $F \leftarrow F \cup F_2$;
(15) **For** each frequent *itemset*, $i_s i_t$, in $F_2$ **do**
(16)     Create the root of a tree, $R_{st}$, and label it by $i_s i_t$;
(17)     ***Constructing_Pattern_Tree***$(R_{st}, \{i \mid i \in F_1, i \succ i_s\}, \varnothing)$;
(18) **Endfor**
(19) **Return** $F$;

Procedure ***Building_Pattern_Tree*** () is almost the same as the homonymous procedure presented in [11] except that DiffNodeset replace Nodesets. The procedure employs the superset equivalence property [11] to pruning the search space. The property is described as follows: Given item $i$ and itemset $P$ ($i \notin P$), if the support of $P$ is equal to the support of $P \cup \{i\}$, the support of $A \cup P$, where $A \cap P = \varnothing \wedge i \notin A$, is equal to the support of $A \cup P \cup \{i\}$. Let's take Figure 3 as an example. If we find the support of *abc* is equal to the support of *ab*, the subtree, whose root is the node registering *c* and representing *abc* in the left of Figure 3, will be pruned in the mining process. *Nd*, *Cad_set*, *ex_frequent_itemsets* are three input parameters. *Nd* stands for the current node in the set-enumeration tree. *Cad_set* are available items that are used to extend Node *Nd*. In fact, *Cad_set* are used to generate child nodes of *Nd*. *FIS_parent* are the frequent itemsets generated on the parent of *Nd*.

Line (4) to (20) check each item in *Cad_set* to find the promoted items and the items that will be used to construct the child nodes of *Nd*. Line (8) builds the DiffNodeset of itemset *P* and line (9) computes *P*'s support. Line (10) and (11) identify a promoted item and inserts it into *Nd.equivalent_items*. An item, *i*, is called promoted if the support of $\{i\} \cup Nd.itemset$ is equal to the support of *Nd.itemset*. Because all information about the frequent itemsets relevant to the promoted items is stored in *Nd*, we don't not need to use the promoted items to further generate the child nodes (actually, subtrees) for discovering frequent itemsets [8]. We call this pruning technique promotion. The pruning efficiency of FIN+ mainly depends on identifying the promoted items. Line (12) to (17) look for all items with which the extension of *Nd.itemset* are frequent. These items are stored in *Next_Cad_ set* for the next procedure, which generates the child nodes of *Nd*. Line (22) to (28) find all frequent itemsets on *Nd*. If *FIS_parent* is null as shown by Line (24), *PSet* is the set of all frequent itemsets on *Nd*. Otherwise, the itemsets, which are generated by *PSet* and *FIS_parent* as Line (27) does, are all frequent itemsets on *Nd*. *FIT_Nd* stores these frequent itemsets for the future procedure of constructing the child nodes of *Nd*. Line (30) to (34) continue to extend the child nodes of *Nd* by recursively calling ***Building_Pattern_Tree*** ().

**Procedure *Constructing_Pattern_Tree*** (*Nd, Cad_set, FIS_parent*)
(1) *Nd.equivalent_items* $\leftarrow \varnothing$;
(2) *Nd.childnodes* $\leftarrow \varnothing$;
(3) *Next_Cad_ set* $\leftarrow \varnothing$;
(4) **For** each $i \in Cad\_set$ **do**
(5)     $X \leftarrow Nd.itemset$;
(6)     $Y \leftarrow (X - X.last\_item) \cup \{i\}$; // $X - X.last\_item$ is the subset by deleting the last item from *X*.
(7)     $P \leftarrow X \cup \{i\}$;
(8)     $P.DiffNodeset \leftarrow X.DiffNodeset \ / \ Y.DiffNodeset$;
(9)     $P.support \leftarrow X.support - \sum_{(E \in P.DiffNodeset)} E.count$;
(10)    **If** $P.support = X.support$ **then**
(11)        $Nd.equivalent\_items \leftarrow Nd.equivalent\_items \cup \{i\}$;
(12)    **Else if** $P.support \geq |DB| \times \xi$, **then**

(13)          Create node $Nd_i$;
(14)          $Nd_i.label \leftarrow i$;
(15)          $Nd_i.itemset \leftarrow P$;
(16)          $Nd.childnodes \leftarrow Nd.childnodes \cup \{Nd_i\}$;
(17)          $Next\_Cad\_set \leftarrow Next\_Cad\_set \cup \{i\}$;
(18)       **Endif**
(19)    **Endif**
(20) **Endfor**
(21) **If** $Nd.equivalent\_items \neq \varnothing$ **then**
(22)    $SS \leftarrow$ the set of all subsets of $Nd.equivalent\_items$;
(23)    $PSet \leftarrow \{A \mid A = Nd.label \cup A', A' \in SS\}$;
(24)    **If** $FIS\_parent = \varnothing$, **then**
(25)       $FIT\_Nd \leftarrow PSet$;
(26)    **Else**
(27)       $FIT\_Nd \leftarrow \{P' \mid P' = P_1 \cup P_2, (P_1 \neq \varnothing \wedge P_1 \in PSet)$ and $(P_2 \neq \varnothing \wedge P_2 \in FIS\_parent\}$;
(28)    **Endif**
(29)    $F \leftarrow F \cup FIT\_Nd$;
(30) **Endif**
(31) **If** $Nd.childnodes \neq \varnothing$ **then**
(32)    **For** each $Nd_i \in Nd.childnodes$ **do**
(33)       ***Constructing_Pattern_Tree***$(Nd_i, \{j \mid j \in Next\_Cad\_set, j \succ i\}, FIT\_Nd)$;
(34)    **Endfor**
(35) **Else** *return*;
(36) **Endif**

## 5. EXPERIMENTAL EVALUATION

In the experiments, we choose FIN [10], PrePost [13], FP-growth* [15], and Eclat_g as the comparison algorithms. FIN and PrePost are two best algorithms which employ node sets to represent itemsets. FP-growth* is the state-of-the-art algorithm among all FP-tree-based pattern growth methods [15], and is the winner of FIMI 2003. Eclat_g (abbreviation for eclat_goethals), is one of the best vertical mining algorithms [13]. FIN+, FIN, and PrePost were all implemented in C++. The implementation of FP-growth* and Eclat_g in C++ were downloaded from http://fimi.ua.ac.be/src/ and http://www.adrem.ua.ac.be/~goethals/software/ respectively. All the experiments are performed on a computer with 14G memory and Intel Xeon @2.0GHZ processor. The operating system is Windows Server 2003 Standard x64 Edition. Note that all these algorithms discover the same frequent itemsets, which confirms the result generated by any algorithms in our experiments is correct and complete.

### 5.1. Datasets

To evaluate FIN+, we used four real datasets and one synthetic dataset, which were often used in previous study of frequent itemset mining, for the performance tests. These real datasets were downloaded from FIMI repository (http://fimi.ua.ac.be). The chess dataset are originated from some game steps. The pumsb dataset contains census data. The kosarak dataset contains click-stream data of an on-line news portal. The mushroom dataset contains characteristics of various kinds of mushrooms. The T10I4D100K dataset is a synthetic dataset and was generated by the IBM generator (http://www.almaden.ibm.com/cs/quest/syndata.html). To generate T10I4D100K, the average transaction size and average maximal potentially frequent itemset size are set to 10 and 4, respectively while the number of transactions in the dataset and different items used in the dataset are set to 100K and 1K, respectively.

Table 2 shows the characteristics of these datasets, where shows the average transaction length (denoted by #Avg.Length), the number of items (denoted by #Items) and the number of transactions (denoted by #Trans) in each dataset. Note that, these real datasets are usually very dense. For example, when the minimum support is set to 5%, the number of frequent itemsets discovered from the mushroom dataset is more than 3 millions. The synthetic datasets generated by the IBM generator mimic the transactions in a retailing environment. Therefore, the synthetic datasets are usual much sparser when compared to the real sets. For example, even if the minimum support is set to as low as 0.1%, the number of frequent itemsets discovered from the T10I4D100K dataset is still less than 30k.

**Table 2. The summary of the used datasets**

| Dataset | #Avg. Length | #Items | #Trans |
|---|---|---|---|
| chess | 37 | 75 | 3,196 |
| pumsb | 74 | 2,113 | 49,046 |
| kosarak | 8 | 41,270 | 990,002 |
| mushroom | 23 | 119 | 8,124 |
| T10I4D100K | 10 | 949 | 98,487 |

## 5.2. DiffNodesets vs. Nodesets

Our first experiment is to compare the advantages of DiffNodesets versus Nodesets in terms of the sizes. We conduct experiment on all real and synthetic datasets. Table 4 shows the average cardinality of the DiffNodesets and Nodesets for frequent itemsets of various lengths on different dataset, for a given minimum support. Note that, in Table 4, Reduction Ration means the value of the average DiffNodeset size divided by the average Nodeset size. We find DiffNodesets are smaller than Nodesets for all datasets. Specially, on the dense datasets, such as connect and pumsb, DiffNodesets are smaller by one to two orders of magnitude than Nodesets.

**Table 3. The summary of the used datasets**

| Dataset | Min-Sup ($\xi$) | #Avg. DiffNodesets | #Avg. Nodesets | Reduction Ration |
|---|---|---|---|---|
| chess | 15% | 0.5 | 367.7 | 735 |
| pumsb | 50% | 12 | 431 | 36 |
| kosarak | 0.2% | 282 | 2091 | 7.4 |
| mushroom | 5% | 98 | 134 | 1.4 |
| T10I4D100K | 0.1% | 81 | 105 | 1.3 |

## 5.3. Runtime Comparison

Figure 4, 5, 6, 7, and 8 show the runtime comparison of dFIN against FIN, PrePost, FP-growth*, and Eclat_g. In these figures, the X and Y axes stand for minimum support and running time respectively. It should be noticed that runtime here means the total execution time, which is the period between input and output. To thoroughly evaluation the performance of runtime, we conduct extensive experiments spanning all the real and synthetic datasets mentioned above with various values of minimum support.

Figure 4 gives the result for running the five algorithms on the chess dataset. We find that dFIN is the fastest one among all algorithms for all minimum supports. Although PrePost run faster than FP-growth* and Eclat_g, the difference among the three algorithms is not distinguished. FIN performs worst and is about one or two orders of magnitude slower than dFIN. The reason is that the average size of DiffNodesets is about two orders of magnitude smaller than that of Nodesets on the chess dataset as shown in Table 3.

Figure 5 shows the result for running all five algorithms on the pumsb dataset. We observer that dFIN is still the fastest one among all algorithms for all minimum supports. PrePost run faster than FP-growth* for high minimum supports while FP-growth* run faster than PrePost for low minimum supports. Eclat_g runs slowest among all algorithms for all minimum supports. Although FIN is not the worst one, it is still slower by one order of magnitude than dFIN. The reason can be explained by the result in Table 3 again. As shown in Table 3, the average size of DiffNodesets is about one order of magnitude smaller than that of Nodesets on the pumsb dataset.

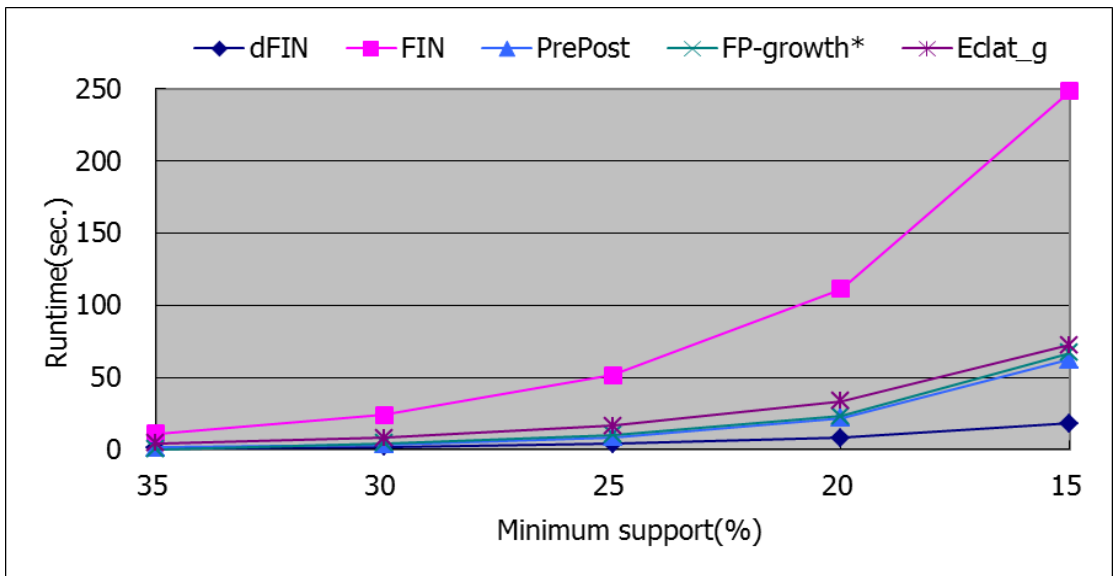

Figure 4. Runtime on the chess dataset

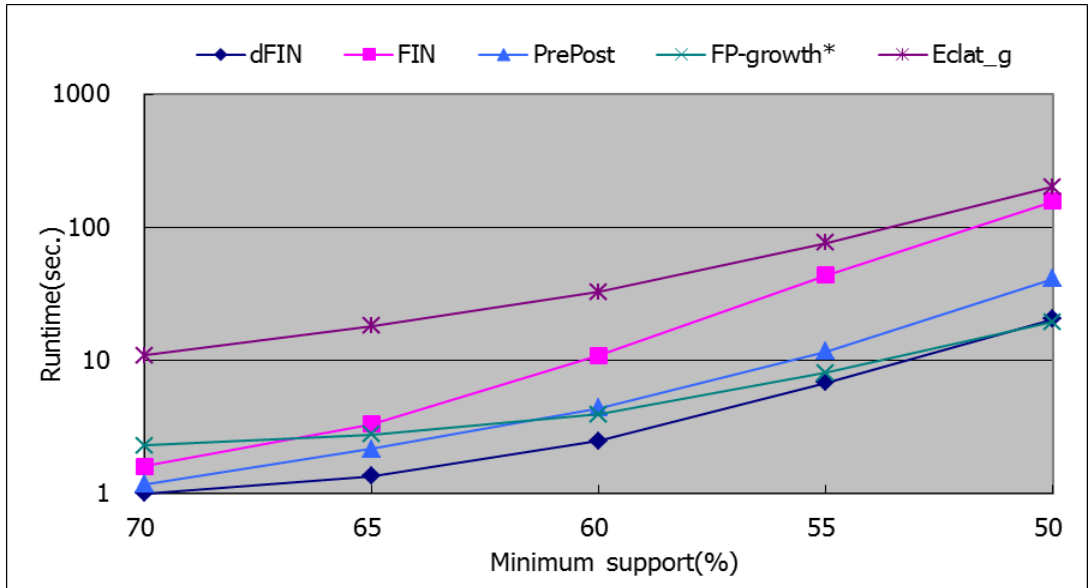

Figure 5. Runtime on the pumsb dataset

Figure 6, 7, and 8 gives the result for running all five algorithms on dataset kosarak, mushroom, and T10I4D100K. In the figures, Eclat_g shows the worst performance on each dataset for both high and low minimum supports. On the sparse synthetic dataset, T10I4D100K, FP-growth* has performance similar to dFIN, FIN, and PrePost. However, on other two real dataset, FP-growth* is apparently slower than dFIN, FIN, and PrePost. Although dFIN performs a litter better than FIN and PrePost on all three dataset, the difference between them is negligible. As shown in Table 3, compare with Nodesets, the average size of DiffNodesets does not decrease too much on dataset kosarak, mushroom, and T10I4D100K. This explains the reason that the advantage of dFIN over FIN on these three datasets is not as distinguished as that on dataset chess and pumsb.

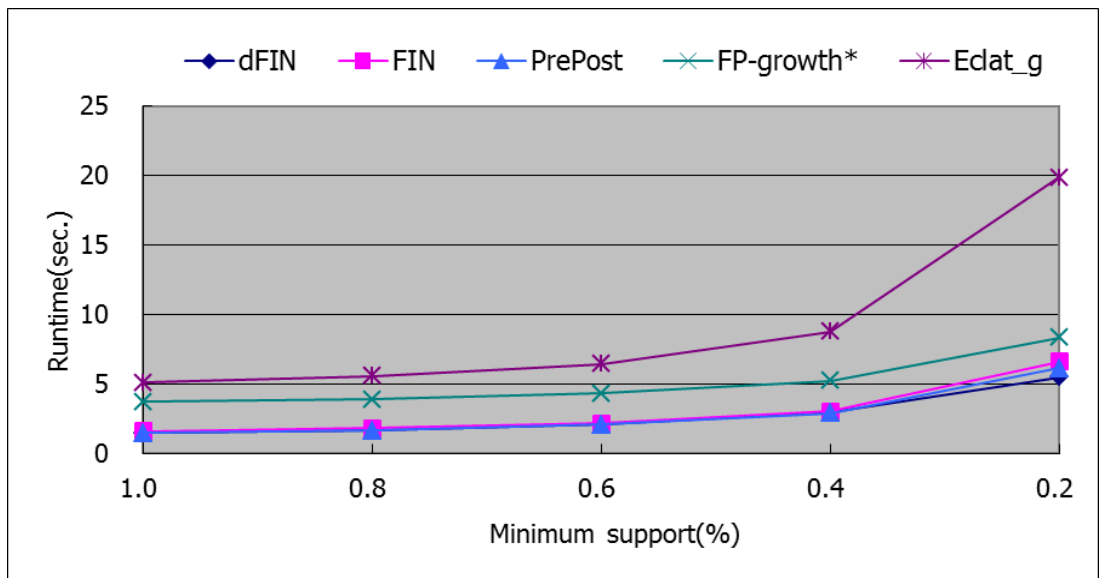

Figure 6. Running time on the kosarak dataset

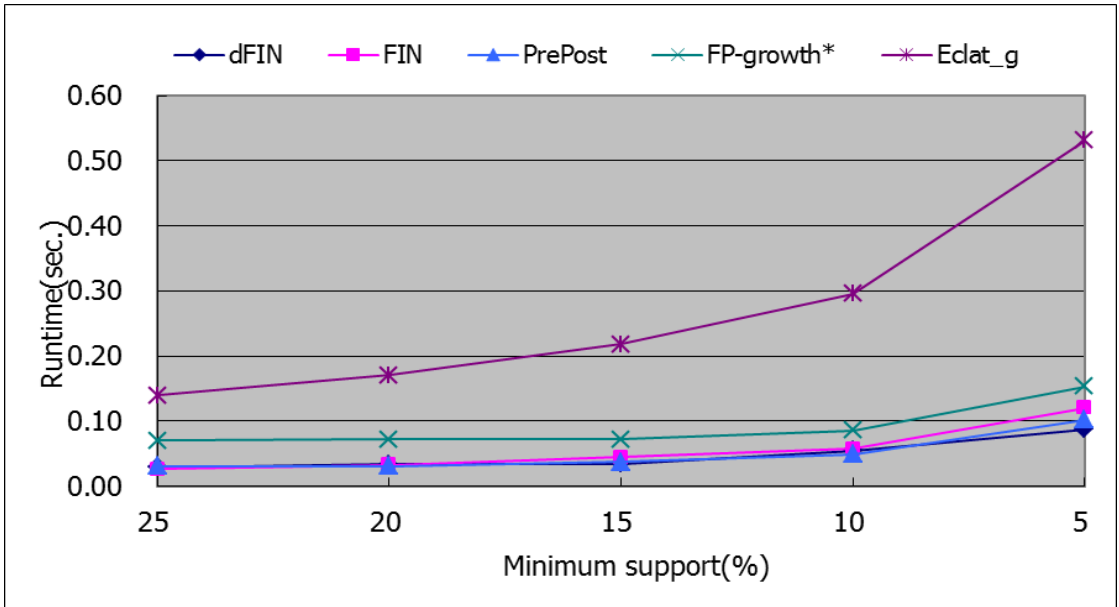

Figure 7. Running time on the mushroom dataset

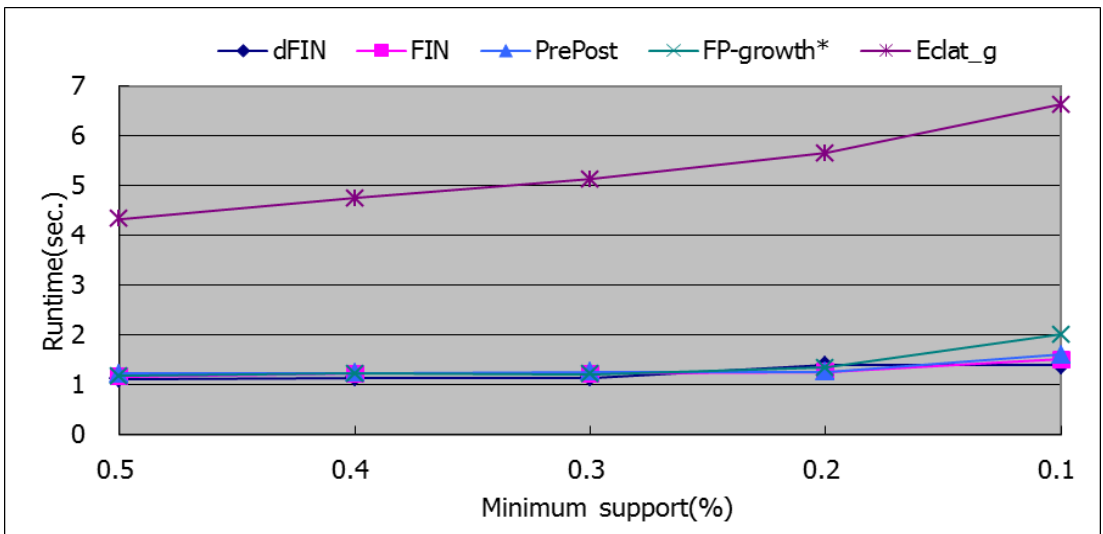

Figure 8. Running time on the T10I4D100K dataset

## 5.3. Memory Consumption

Figure 9 to 13 shows the peak memory consumption of all five algorithms on the five datasets.

In Figure 9, dFIN and FIN consume almost the same memory on the chess dataset. In the figure, Eclat_g uses the lowest amount of memory for all minimum supports. FP-growth* also consumes far less memory than the other algorithms except Eclat_g. As the minimum support becomes small, the peak memory consumption of dFIN, FIN, and PrePost increases faster than Eclat_g and FP-growth*.

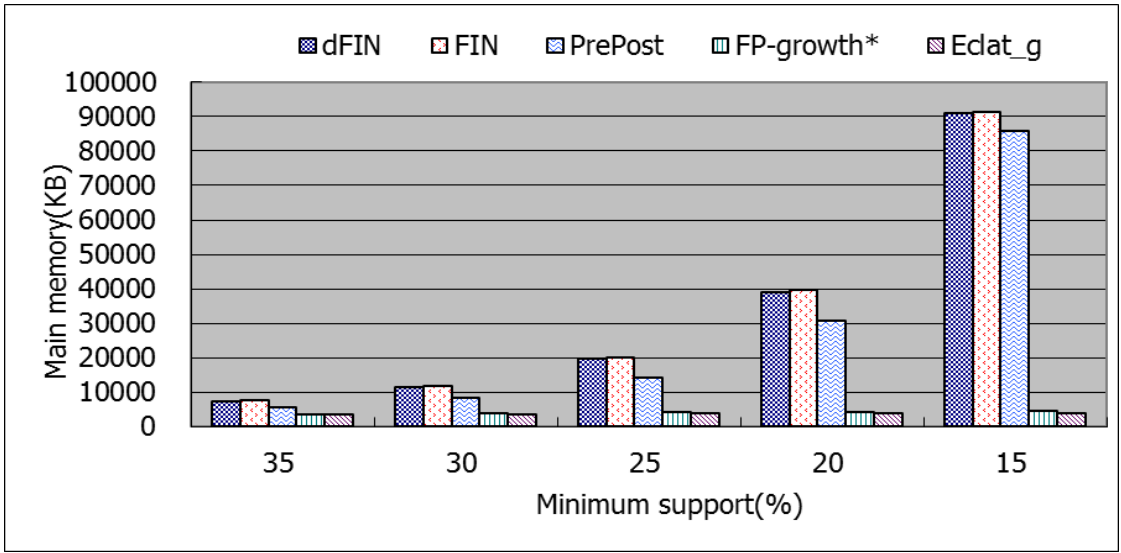

Figure 9. Memory consumption on the chess dataset

Figure 10 shows the result on the pumsb dataset. In the figure, dFIN has performance similar to FP-growth*. Both of them use less memory than FIN, PrePost, and Eclat_g. When the minimum support is high, FIN and PrePos consume less memory than Eclat_g. However, Eclat_g use less memory than FIN and PrePos when the minimum support is low.

Figure 11 shows the result on the kosarak dataset. In the figure, dFIN and FIN consume almost the same memory. Both of them use a little more memory than FP-growth*, but less than PrePost. Eclat_g uses the lowest amount of memory when the minimum support is lowest. However, Eclat_g consumes more memory than the other four algorithms when the minimum support is high.

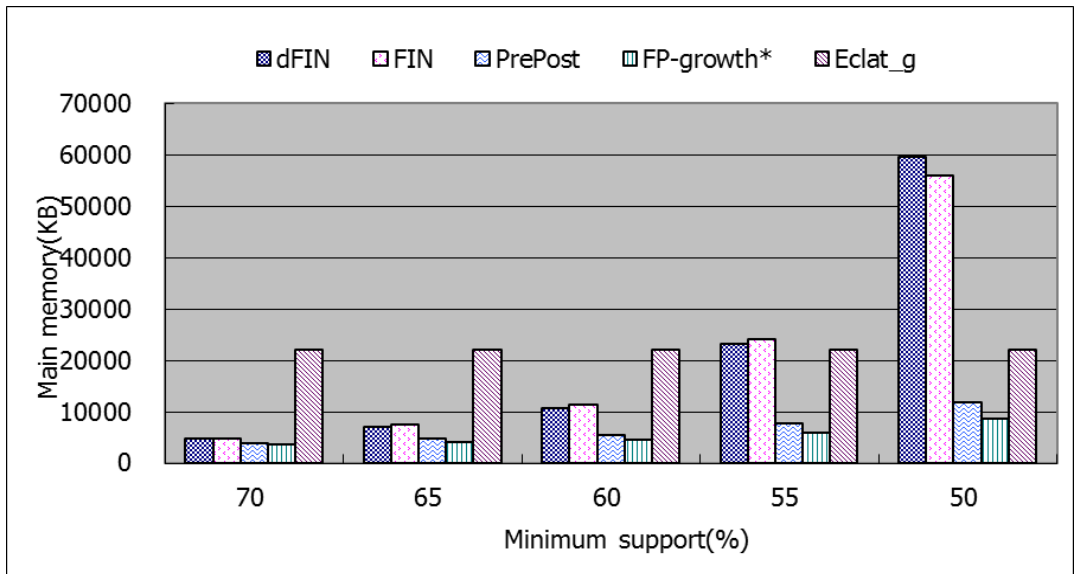

Figure 10. Memory consumption on the pumsb dataset

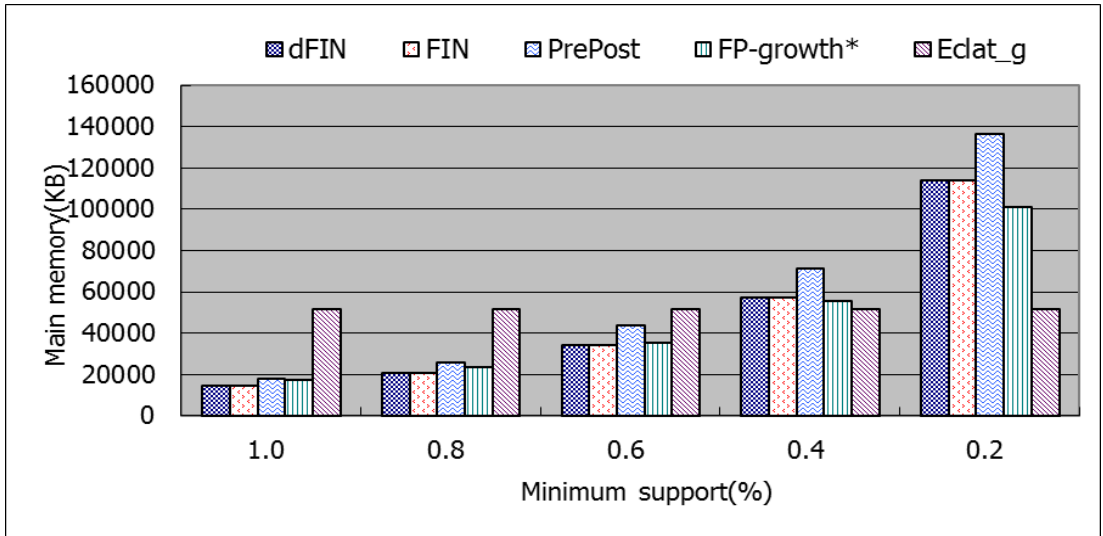

Figure 11. Memory consumption on the kosarak dataset

Figure 12 shows the result on the mushroom dataset. In the figure, dFIN, FIN, and PrePost consume almost the same memory. They all use a little more memory than FP-growth*. Once again, Eclat_g consumes the lowest amount of memory for the lowest minimum support while it consumes more memory than other algorithm when the minimum support is high.

The result on the T10I4D100K dataset is shown by Figure 13. In the figure, Eclat_g consumes the lowest amount of memory for each minimum support while PrePost consumes the highest amount of memory. dFIN and FIN consume almost the same memory. Although FP-growth* perform better than dFIN and FIN. The difference among them is small.

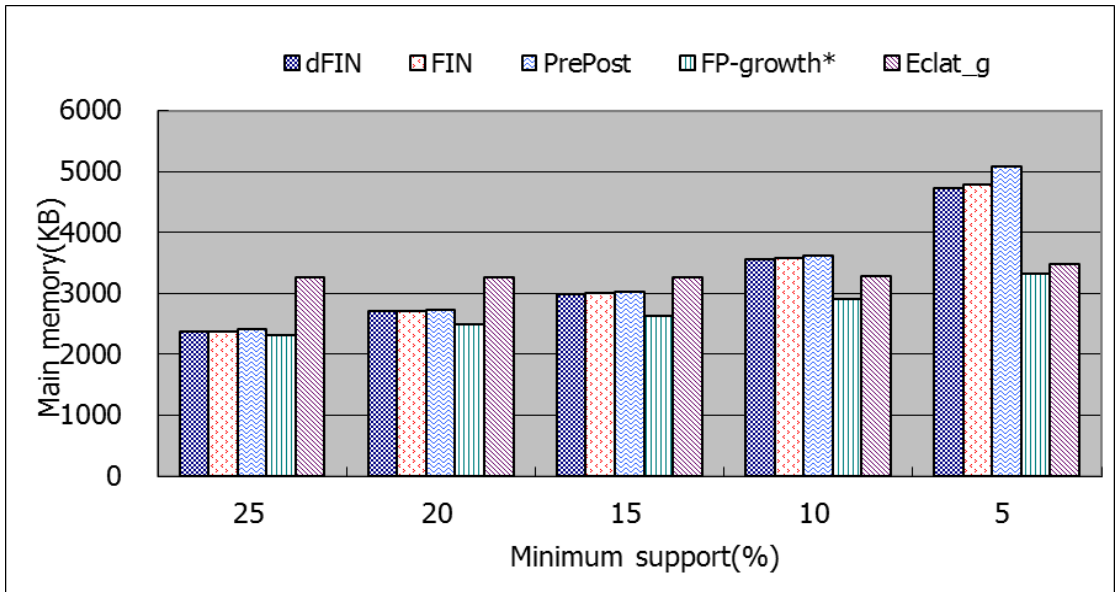

Figure 12. Memory consumption on the mushroom dataset

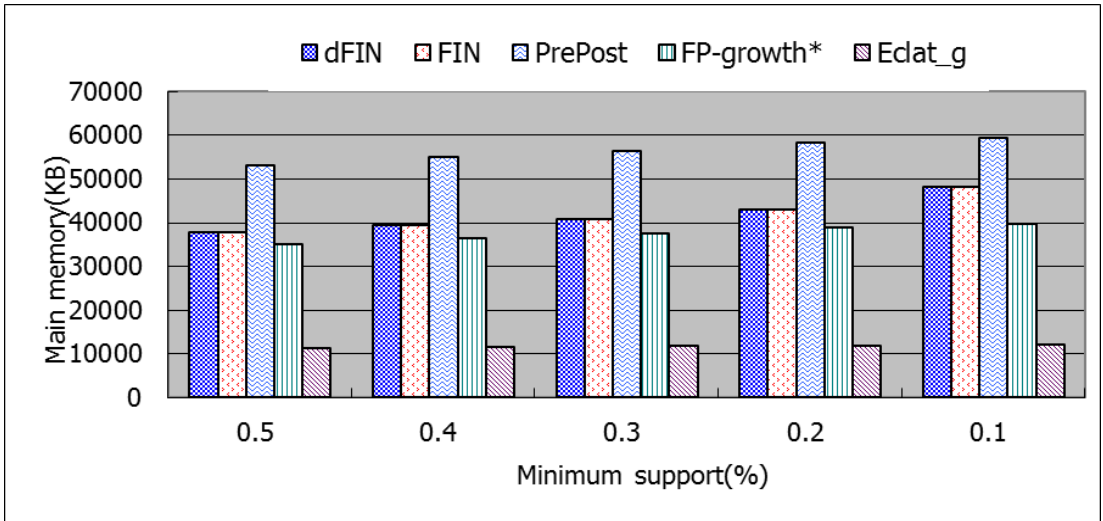

Figure 13. Memory consumption on the T10I4D100K dataset

From these figures, we find that dFIN and FIN consume almost the same memory and perform worse than PrePost. This can explained as follows. The main component of memory consumption is the original PPC-tree or FP-tree. Since a node of PPC-tree contains more information (*pre-order* and *post-order* number) than a node of FP-tree, the PPC-tree of a dataset is a little bigger than its FP-tree. In addition, dFIN needs to generate all frequent 1-itemsets with associated DiffNodesets or Nodesets meanwhile maintaining a PPC-tree. This needs large memories when there are a lot of frequent 1-itemsets.

In addition, we also notice that when the memory used by Eclat_g changes little when minimum support changes. We guess the reason is the Eclat_g adopts some techniques to ensure the constancy of memory consumption.

## 6. CONCLUSIONS

In this paper, we present a novel structure called DiffNodeset to facilitate the process of mining frequent itemsets. Based on DiffNodesets, an algorithm named FIN+ is proposed to fast find all frequent itemsets in databases. Compared with Nodeset, the key advantage of DiffNodeset lies in that its size much smaller. This makes DiffNodeset more suitable for mining frequent itemsets. The extensive experiments show that DiffNodeset is favorable. FIN+ proves to be state-of-the-art since it always runs fastest on all datasets with different minimum supports when compared with previous leading algorithms.

As future work, first we will investigate how to integrate DiffNodesets with other techniques, such as Principle of Inclusion–Exclusion [20] and Linear Prefix tree [23], to further promote the mining efficiency. Second, we will explore how to apply DiffNodesets to mine maximal frequent itemsets [3, 4], closed frequent itemsets [18, 29], frequent disjunctive closed itemsets [28], and Top-Rank-k frequent patterns [8]. Third, we will further extend DiffNodesets to mine frequent itemsets over data streams [6, 19, 30]. Finally, it is an interesting work to study how to use DiffNodesets to mine frequent itemsets under parallel/distributed architecture [27].

## 7. ACKNOWLEDGEMENTS

This work is partially supported by Project 61170091 supported by National Natural Science Foundation of China. We are also grateful to the anonymous reviewers for their helpful comments.